\documentstyle[11pt]{article}
\begin{document}
\renewcommand{\thepage}{ }
\begin{titlepage}
\title{
\hfill
\vspace{1.5cm}
{\center Glassy Behavior in the Ferromagnetic Ising Model on a Cayley Tree}
}
\author{
R. M\'elin$^{(1)}$, J.C. Angl\`es d'Auriac$^{(1)}$,
P. Chandra$^{(2)}$
and B. Dou\c{c}ot$^{(1)}$\\
{}\\
{$^{(1)}$CRTBT-CNRS, 38042 Grenoble BP 166X c\'edex France}\\
{$^{(2)}$NEC Research Institute, 4 Independence Way,
08540 Princeton NJ, USA}}
\date{}
\maketitle
\begin{abstract}
\normalsize
We present a detailed study of the
nearest-neighbor ferromagnetic Ising model on
a Cayley tree.  In the limit of zero field,
the system displays glassy behavior below a crossover
temperature, $T_g$, that scales inversely
with the logarithm of the number of generations;
thus $T_g$ is inversely proportional to
the logarithm of the logarithm of the number of
sites.
Non-Gaussian magnetization distributions are
observed for $T < T_g$, reminiscent of that
associated with the central spin of
the Edwards-Anderson model on
the same tree; furthermore a dynamical
study indicates metastability, long relaxation
times and ageing consistent with the
development of glassy behavior for a finite
but macroscopic number of sites.
\end{abstract}
\end{titlepage}

\newpage
\renewcommand{\thepage}{\arabic{page}}
\setcounter{page}{1}
\baselineskip=17pt plus 0.2pt minus 0.1pt

\section{Introduction}

Recursive structures like the Bethe lattice and the Cayley
tree provide a pedagogical enviroment for the study of physical
problems; in this setting they can be treated with a direct
analytic approach without resorting to approximate methods.\cite{thorpe}
The Bethe lattice, an infinite Cayley tree, is a connected
dendritic structure with constant coordination, $z$, and {\sl no}
loops, as displayed for $z=3$ in Figure \ref{fig1a}.  Strictly speaking
it is a {\sl pseudo-lattice} since it cannot be embedded
in any {\sl real} finite-dimensional lattice; indeed it is
often regarded as an infinite-dimensional structure since the number
of sites accessible in $N$ steps from a given site ($\sim N^d$ for
a $d$-dimensional lattice) increases exponentially with $N$.  Thus
the Bethe lattice provides a setting where mean-field treatments
can become exact.  This property was first discussed by Domb
who showed that the Bethe-Peierls (BP) approximation to the
nearest-neighbor (nn)
ferromagnetic (FM) Ising problem, with
$H = -J \sum_{(ij)} \sigma_i \sigma_j$ where $J > 0$,
$\sigma_i = \pm 1$ and $(ij)$ indicates a nn sum,
is exact on this structure;\cite{domb}
its solution is identical to that of the infinite-range
FM Ising model.\cite{baxter}  Similarly Thouless, Anderson and Palmer
studied the infinite-range Sherrington-Kirkpatrick (SK) model
of spin glasses on the Bethe lattice using a mean-field
technique;\cite{tap} they were able to recover the key results\cite{sk} of
Sherrington and Kirkpatrick (SK) without using the replica method.
More recently there have been several studies of the SK model
on the Bethe lattice, particularly in finite fields.\cite{1,2,3,4,5,6,7,8,9}
In general the study of a variety of problems on this recursive
structure has helped to develop our understanding of
diverse physical phenomena including self-avoiding polymers,\cite{polymers}
random resistor networks\cite{stinchcombe} and
percolation.\cite{rammal}

Like the Bethe lattice, a Cayley tree is a connected structure
with fixed coordination number and no loops; however it has a {\sl
finite}
number of generations (cf. Figure \ref{fig1a})
and hence sites that are dominated
by the boundary.  More specifically, the total number of sites
in a Cayley tree of $n$ generations  with coordination $z$ is
\begin{equation}
N = 1 + z + z(z-1) + ...z(z-1)^{n-1} = \frac{(z(z-1)^n - 2)}{(z-2)}
\end{equation}
and the number of surface atoms  is
\begin{equation}
N_s = z(z-1)^{n-1}
\end{equation}
so that for large $n$
\begin{equation}
\frac{N_s}{N} \approx \frac{(z-2)}{(z-1)}
\end{equation}
in contrast to the situation in ``real'' lattices $\left(\frac{N_s}{N} \sim
N^{-\frac{1}{d}}\right)$.  Thus the ``interior'' of a Cayley tree, in the limit
of a large number of generations, contains an arbitrarily small
fraction
of its total number of sites, and the boundary
plays a key role in any problem studied on this
graph. In particular, the Bethe-Peierls
transition for the FM Ising model on a Cayley tree
occurs {\sl only} for its central spin; despite its finite moment,
the total spontaneous magnetization
of {\sl all} the spins remains zero.\cite{10,11,12,13}
In a nutshell this occurs because, at zero field and low temperatures,
very large domains of flipped spins can nucleate from
the boundaries; the resulting finite-size glassiness
is the subject of this Paper.

The recursive structure of the Cayley tree permits
a detailed analysis of the single-site magnetization
distribution as a function of field and generation.
In doing so, we find that for fields $h < h_{c0}$,
where the crossover field $h_{c0}$ decreases exponentially with the number
of generations, there is  a temperature-scale
$T_g$ below which well-defined, large domains
of flipped spins exist.  For $T < T_g$ the magnetization
distribution becomes non-Gaussian, reminiscent of
that associated with the central spin of the magnetized spin glass
phase of the $\pm J$ model on the Cayley tree.
The cross-over temperature, $T_g$, scales inversely
with the logarithm of the number of generations
of the Cayley tree so that the ``finite-size'' glassiness
persists to very large system sizes for a macroscopic number of sites.

We therefore have a short-ranged periodic spin model that has
a "glass" cross-over temperature that decreases very slowly
with increasing system size; more specifically it is inversely
proportional to the logarithm of the logarithm of the number
of sites.  We characterize it using a combination of analytic and
numerical techniques, always retaining  open boundary conditions.
First we study the magnetization for different
thermodynamic limits emphasizing the crucial role of
the ratio of surface/bulk sites as $n \rightarrow \infty$.
We recover  the Bethe-Peierls result if
this ratio goes to zero; otherwise, for vanishing applied fields,
there is a cross-over to a glassy phase characterized by
well-defined clusters of flipped spins.  We find in
that  for $h < h_{co}$,
the single-site magnetization distribution
becomes non-Gaussian for $T <T_g$
similar to that of spin glass models residing on
the same structure; however it recovers its Gaussian
character with increasing field. For $T < T_g$
the largest barriers associated with developing
broken bonds in these domains scale with the number of
cluster sites; we thus refer to this low-temperature state
as a finite-size glass.  A dynamical study of this system,
performed numerically, indicates
the presence of metastable states and long relaxation times
at low temperatures.
The autocorrelations for $T < T_g$ are determined after a waiting
time, and indicate ageing effects; the variation of
$\chi_1'$ and $\chi_3'$ with temperature also agree with
the presence of glassiness.  As expected, the Edwards-Anderson
susceptibility of the entire tree has a maximum which develops
slowly with system-size; no divergence is observed.
Finally we find that for this system
the density of Lee-Yang zeroes in the presence
of a complex field
is very high in the vicinity of the real
axis below a particular temperature-scale; we identify it as
the crossover temperature $T_g$.
We end with a summary of
our results and plans for future work.

\section{The Different Thermodynamic Limits on the Cayley Tree}
\subsection{Warm-Up: The Bethe-Peierls
Transition of the Central Spin}

One way to take the thermodynamic limit on
an $n$-generation Cayley tree
is to look {\sl solely}  at the properties of its
central spin, and then to take
the limit $n \rightarrow + \infty$.
As first pointed out in \cite{gujrati},
the behavior of the central spin is then
characteristic of an infinite-dimensional lattice; more specifically
it displays
a mean-field transition.
In other words, the Bethe-Peierls approximation
becomes exact on a
finite Cayley tree if and only if
one considers solely the
properties of its central spin and ignores its surface.
For recent results
in this field, we refer the reader to reference \cite{polymers}
In appendix A we give the
calculation
of the recursion relation
for the partition function $Z_n(\beta,H,H_n)$ of an $n$-generation
tree with
coordination $z$;
here $\beta$
is the inverse temperature and
$H$ and $H_n$ are the magnetic fields  acting on the spins of
generations $0$ to $n-1$ and $n$ respectively.
The resulting recursion relation for $Z_n(\beta,H,H_n)$ is
\begin{equation}
Z_n(\beta,h,h_n) = \left(4 \cosh^{2}{(\beta J)}
\cosh^{2}{(\beta h_n)}
\right)^{\frac{(z-1)^{n}}{2}}
Z_{n-1}(\beta,h,h+T.h_n)
,
\end{equation}
where the transformation of the magnetic field is
\begin{equation}
T.h = \frac{1}{2 \beta}
\ln{\frac{\cosh{\beta(J+h)}}{\cosh{\beta(J-h)}}}
\label{eq17}
{}.
\end{equation}
We consider the  special case  of a
field $\epsilon$ applied at the surface of a tree,
and ask whether it is amplified in the bulk
as determined by the recursion relation
(\ref{eq17}). This condition defines
the bulk critical temperature
$\beta_c$ by the expression $(z-1) \tanh{\beta_c J} = 1$.
If $\beta<\beta_c$, the magnetization of
the central spin is zero; however if
$\beta > \beta_c$, there is a broken symmetry
for the central spin in the
thermodynamic limit. As discussed in \cite{baxter},
the critical behavior of the central
spin in the thermodynamic limit
is identical to that of an infinite-range
ferromagnet
where $\beta = 1/2$ and
$\delta = 3$.

\subsection{Beyond the Bethe Peierls Regime:
Different Thermodynamic Limits}
We now wish to look at the transition,
not only of the central spin,
but of the entire tree. First
we consider only {\it half-space-trees},
that is trees such that the coordination associated
with the ancestor is $z-1$
and not $z$. In what follows, we denote a  half-space-tree of
$n$ generations
as a $n$-half-space-tree; one
is pictured on figure
\ref{fig1}. We shall label  the generations
so that the ancestor is
at generation $n$ and the leaves are at generation 1.
We are interested in the magnetic properties of the
spins in generations $n-m$ to $n$
in the limit $n \rightarrow + \infty$ where,
of course, $m$ is a function
of $n$; more specifically we want to classify the different
regimes as a function
of $m(n)$ in an external uniform field $h$.
In order to obtain
the magnetization of the generations $n-m$ to $n$,
we must
apply a source magnetic field $\lambda$
to these generations; then
we differentiate the partition function
with respect to $\lambda$
in the limit $\lambda \rightarrow 0$ to obtain
\begin{equation}
\langle M(n,m,h) \rangle =
\frac{\partial}{\partial(\beta \lambda)}
\ln{Z(n,m,h,\lambda=0)}
{}.
\end{equation}
Details of the calculation of
the partition function are given
in appendix B, and we obtain
\begin{equation}
\langle M(n,m,h) \rangle =
\sum_{i=n-m}^{n-1}
(z-1)^{n-i}  \frac{\sinh{\beta h_i}
\cosh{\beta h_i}}
{\cosh^{2}{\beta J} + \sinh^{2}{\beta h_i}}
\frac{d h_i}{d \lambda}(\lambda=0) +
\tanh{\beta h_n}
\frac{d h_n}{d \lambda}(\lambda = 0)
\label{eq18}
{}.
\end{equation}
The last term is the contribution of the
ancestor to the average magnetization;
$h_i$ is the total field at generation $i$
which is the sum of the external, the source
and the recursive (of (\ref{eq17}) fields ($h$,$\lambda$ and $h^{ind}$)
respectively.

We do not treat the iteration of $h^{ind}$ exactly,
but approximate it as described below.
 From the iteration
(\ref{eq17}) we can deduce the shape of
$h^{ind}_{i+1}$ as a function of $h_i$,
which is plotted on figure
\ref{fig1bis}. If $\beta < \beta_c$,
the slope at the origin
is less than unity, whereas it is larger
than unity for $\beta > \beta_c$.
Moreover for $\beta > \beta_c$ there is
one non-trivial fixed point,
for which $h_{i+1}^{ind} = h_i = h^{*}$
which depends only on the temperature.
This behavior suggests that the iteration
can be approximated by
linearizing $h_{i+1}(h_i)$ in the vicinity
of $h_i=0$ and of $h_i=h^{*}$.
More precisely, we deduce from (\ref{eq17}) that
\begin{equation}
\frac{d h^{ind}_{i+1}}{d h_n} = (z-1)
\frac{\sinh{\beta J}
\cosh{\beta J}}{\cosh^{2}{\beta J} +
\sinh^{2}{\beta h_i}}.
\end{equation}
and define new variables, $\eta_1$ and $\eta_2$,
such that
\begin{eqnarray}
\frac{d h^{ind}_{i+1}}{d h_i}(h_i=0)
&=& (z-1) \tanh{\beta J}
\equiv 1 + \eta_1\\
\frac{d h^{ind}_{i+1}}{d h_i}(h_i=h^{*})
&\equiv& 1 - \eta_2
,
\end{eqnarray}
where $\eta_1 \in [0,z-2]$ and $\eta_2 \in [0,1]$
and we note that
$\eta_1$ and $\eta_2$ depend only on the temperature.
We then express the complete recursion by
\begin{equation}
h^{ind}_{i+1} = (1  + \eta_1) h_i
,
\end{equation}
or
\begin{equation}
h_{i+1}^{ind} = \eta_2 h^{*} + (1-\eta_2) h_i
{}.
\end{equation}
The first linearization corresponds to $h_i \in [0,h_c]$
and the second one to $h_i \in [h_c,h^{*}]$, where
\begin{equation}
h_c = \frac{\eta_2}{\eta_1 + \eta_2} h^{*}
{}.
\end{equation}
If $\lambda=0$, iteration of the total
magnetic field $h_n$ leads to different results
depending on the
relative
magnitude of $i$ (generation) compared with
\begin{equation}
n_c(h) = \left[ \frac{1}{\ln{(1+\eta_1)}}
\ln{
\left( \frac{\eta_1 \eta_2}{\eta_1+\eta_2}
\frac{h^{*}}{h} + 1
\right)} \right]
\label{n_c}
\end{equation}
where $[]$ denotes the integer part;
for $i \le n_c(h)$
\begin{equation}
h_i = \frac{h}{\eta_1} \left(
(1 + \eta_1)^{i+1} -1 \right)
\end{equation}
and if $i \ge n_c(h)$.
\begin{equation}
h_i = h^{*} + \frac{h}{\eta_2} +
\left( h_{n_c} - h^{*} - \frac{h}{\eta_2}
\right) ( 1 - \eta_2)^{i-n_c}.
\end{equation}
There are therefore three regimes which
we can study corresponding to the
application of large (I), intermediate (II)
and vanishing (III) fields on the finite
Cayley tree; more specifically
they correspondand to the conditions
$0 \le n_c \le n-m$ (I),
$n-m +1\le n_c \le n$ (II)
and $n_c \ge n$ (III) respectively
where $n_c$ is defined in equation (\ref{n_c}).

\subsubsection{Regime I: $0 \le n_c \le n-m$}

We
now determine the average
magnetization as  given by equation
(\ref{eq18}) in Regime I. Using the approximation discussed
in the previous section, we write
\begin{equation}
\frac{d h_i}{d \lambda}(\lambda=0) = \frac{1}{\eta_2}
(1 - (1-\eta_2)^{i-n+m+1})
{}.
\end{equation}
If $0 \le h_i \le h_c$ or $i < n_c$,
we can approximate
\begin{equation}
\frac{\sinh{\beta h_i} \cosh{\beta h_i}}
{\cosh^{2}{\beta J} + \sinh^{2}{\beta h_i}}
\simeq a_1 h_i
\label{eq19}
,
\end{equation}
where
\begin{equation}
a_1 = \frac{\beta}{\cosh^{2}{\beta J}}
\end{equation}
and if $h_i \ge h_c$ or $n \ge n_c$, we use
\begin{equation}
\frac{\sinh{\beta h_i} \cosh{\beta h_i}}
{\cosh^{2}{\beta J} + \sinh^{2}{\beta h_i}}
\simeq a_2(h_i - h^{*}) + b_2
\label{eq20}
,
\end{equation}
where
\begin{eqnarray}
a_2 &=& \frac{\cosh^{2}{\beta J}
\cosh^{2}{\beta h^{*}}
+ \sinh^{2}{\beta J} \sinh^{2}
{\beta h^{*}}}
{ (\cosh^{2}{\beta J} + \sinh^{2}
{\beta h^{*}})^{2}}\\
b_2 &=& \beta \frac{
\sinh{\beta h^{*}} \cosh{\beta h^{*}}}
{\cosh^{2}{\beta J} + \sinh^{2}{\beta h^{*}}}
{}.
\end{eqnarray}
It is now straightforward
to insert these expressions
into the equation (\ref{eq18}) for
the average magnetization. Since we normalize
by the number of sites
\begin{equation}
N_m = 1 + (z-1) + ... + (z-1)^{m} =
\frac{(z-1)^{m+1} - 1}{z-2}
\label{eq22}
,
\end{equation}
it is reasonable to neglect the contribution
of the ancestor.
We obtain
\begin{eqnarray}
&&\langle M(n,m,h) \rangle = \frac{1}{\eta_2}
\left[ (b_2 + a_2 \frac{h}{\eta_2}) N_m \right.\\
\nonumber
&+& \left. (z-1)^{m}
\frac{1-((1-\eta_2)/(z-1))^{m}}{1-((1-\eta_2)/(z-1))}
\left(-(b_2+a_2 \frac{h}{\eta_2})
(1-\eta_2) + a_2(h_{n_c}-h_{\infty})
(1 - \eta_2)^{n-m-n_c} \right) \right.\\
\nonumber
&-& \left.
a_2 (h_{n_c}-h_{\infty}) (z-1)^{m}
(1-\eta_2)^{n-m-n_c+1}
\frac{1-((1-\eta_2)^{2}/(z-1))^{m}}
{1-((1-\eta_2)^{2}/(z-1))}
\right]
,
\end{eqnarray}
where $h_{\infty} = h^{*} + h/\eta_2$.
We can now discuss the
different thermodynamic limits.
First we note that if $n \rightarrow \infty$,
$m \rightarrow \infty$ and
$n-m \rightarrow \infty$,
the thermodynamic limit of the
normalized magnetization reads
\begin{equation}
\lim \frac{\langle M(n,m,h) \rangle}{N_m}
= \left(b_2 + a_2 \frac{h}{\eta_2}
\right) \frac{z-1}
{z-2+\eta_2}
{}.
\end{equation}
However if we take a {\sl different} thermodynamic limit,
with $n-m \rightarrow b$,
where $b$ is a constant thickness boundary,
we obtain
\begin{equation}
\lim \frac{\langle M(n,m,h) \rangle}{N_m}
= \left(b_2 + a_2 \frac{h}{\eta_2} \right) \frac{z-1}
{z-2+\eta_2} - a_2 (h_{\infty} - h_{n_c})
\frac{(z-1)(z-2) (1-\eta_2)^{b-n_c}}
{(z-2+\eta_2)(z-1-(1-\eta_2)^{2})}
\end{equation}
The absolute value of the corrective term
decreases as the thickness of the boundary increases
since $|1-\eta_2|<1$.

\subsubsection{Regime II: $n - m + 1 \le n_c \le n$}
Again we calculate the average magnetization,
(\ref{eq18}), this time in the regime
$n - m +1\le n_c \le n$
where we use the relations (\ref{eq19})
and (\ref{eq20}) from the previous section. Furthermore
we need to approximate the expression $d h_i/d \lambda$
for $\lambda = 0$.
If $n-m \le i \le n_c$, we have
\begin{equation}
\frac{d h_i}{d \lambda}(\lambda = 0) =
\frac{1}{\eta_1}
\left( (1+\eta_1)^{i-n+m+1}-1 \right)
\label{eq21}
,
\end{equation}
and if $i \ge n_c$,
\begin{equation}
\frac{d h_i}{d \lambda}(\lambda = 0) =
\frac{1}{\eta_2}
+a (1-\eta_2)^{i-n_c}
,
\end{equation}
where
\begin{equation}
a=\frac{1}{\eta_1} (1+\eta_1)^{n_c-n+m+1}
-\frac{1}{\eta_1} - \frac{1}{\eta_2}
{}.
\end{equation}
We have used
\begin{equation}
h_i = h^{*} + \frac{h+\lambda}{\eta_2} +
(1-\eta_2)^{i-n_c}
\left( h_{n_c}(\lambda) - h^{*} -
\frac{h+\lambda}{\eta_2} \right)
,
\end{equation}
with
\begin{equation}
h_{n_c}(\lambda) = h_{n_c}(0) +
\frac{\lambda}{\eta_1}
\left( (1+\eta_1)^{n_c-n+m+1}-1 \right)
{}.
\end{equation}
We insert these expressions into
(\ref{eq18}) and
sum the full geometric series.
We note that the thermodynamic limit can only be taken
with $n-m \rightarrow b$,
where again $b$ is a constant.
We find that the dominant behavior
at small magnetic field $h$ depends
on the temperature.
Let $T'$ be the temperature such as
$\tanh{\beta'} = 1/\sqrt{z-1}$.
Then if $T'<T<T_c$  the dominant term
in the normalized magnetization
is linear in $h$ with corrections
in $h^{\alpha}$,
where $\alpha = \ln{(z-1)}/\ln{(1+\eta_1)} - 1$;
by contrast if
$T<T'$ the leading term is of order $h^{\alpha}$.
We do not write explicitly
the corresponding expressions since they
are tedious and do not contribute
further to the present discussion.

\subsubsection{Regime III: $n_c \ge n$}
In this regime, we approximate the average magnetization
by inserting
(\ref{eq19}) and (\ref{eq21}) into
(\ref{eq18}) which yields
\begin{equation}
\langle M(n,m,h) \rangle =
\sum_{i=n-m}^{n-1}
(z-1)^{n-i} a_1 h \frac{(1+\eta_1)^{i+1} -1}
{\eta_1}
\frac{(1+\eta_1)^{i-n+m+1}-1}{\eta_1}
{}.
\end{equation}
After summing the geometrical series,
we obtain
\begin{eqnarray}
\nonumber
\frac{\langle M(n,m,h) \rangle}{N(m)}
&=&
\frac{a_1 h}{\eta_1^{2}} \frac{z-2}
{1-(z-1)^{-m-1}}
\left[ \frac{1 - (z-1)^{-m}}{z-2} -
\frac{(1 + \eta_1)((1+\eta_1)^{b}+1)}
{z-2-\eta_1}
\left( 1 -\left( \frac{1+\eta_1}{z-1}
\right)^{m} \right) \right.\\
&& \left. + \frac{(1+\eta_1)^{b+2}}
{z-1-(1+\eta_1)^{2}}
\left(1 - \left(\frac{(1+\eta_1)^{2}}
{z-1}\right)^{m} \right) \right]
\end{eqnarray}
We identify the cross-over field $h_{co}$
as $h_{n-1} = h_c$,
that is
\begin{equation}
\label{eq24}
h_{co} = h^{*} \frac{\eta_1 \eta_2}
{(\eta_1 + \eta_2)((1+\eta_1)^{n}-1)}
,
\end{equation}
which is exponentially small in $n$.
If we take the limit of large $n$ and $m$
but finite $n-m$, we can write
\begin{eqnarray}
\nonumber
\frac{\langle M(n,m,h_{co})
\rangle}{N(m)} &=&
h^{*} \frac{a_1 \eta_2}{\eta_1(\eta_1+\eta_2)}
(z-2) (1+\eta_1)^{-n}
\left[ \frac{1}{z-2} - \frac{(1+\eta_1)
((1+\eta_1)^{b}+1)}
{z-2-\eta_1} \left(1-\left( \frac{1+\eta_1}
{z-1} \right)^{m} \right) \right.\\
&& + \left. \frac{(1+\eta_1)^{b+2}}
{z-1-(1+\eta_1)^{2}}
\left( 1 - \left( \frac{(1+\eta_1)^{2}}
{z-1} \right)^{m} \right) \right]
\end{eqnarray}
In the limit $T \rightarrow 0$,
$(1 + \eta_1)/(z-1) = \tanh{\beta}
\rightarrow 1$, so that the only
non-negligible term is
\begin{equation}
\frac{(1+\eta_1)^{-n} (1+\eta_1)^{2 m}}
{(z-1)^{m}}
= (1+\eta_1)^{-b} \left(
\frac{1+\eta_1}{z-1} \right)^{m}
{}.
\end{equation}
Since
\begin{equation}
\ln{\left( \frac{1+\eta_1}{z-1} \right)^{m}}
\simeq - 2 m e^{-2 \beta}
\end{equation}
at small temperature, we obtain the
cross-over temperature scale
$T_g = 2/\ln{m}$.
In the next section we shall rederive
this crossover temperature with another method
and give its significance.

\section{Magnetization Distribution and
Finite-Size Effects}

In this section we study the distribution
of the magnetization which
can be determined from exact recursion
relations.  We find that finite-size
effects are crucial in this analysis,
and we recover the crossover temperature
$T_g$ discussed in the previous section.
We also compute the distribution
of the magnetization in finite field.

\subsection{Magnetization distribution in zero field}
We begin with the case of zero magnetic field.
Given $z-1$ $n$-half-space trees with coordination $z$,
it is straighforward to obtain a $(n+1)$-half
space tree with the same coordination.
One just has to add a common ancestor and to link it
ancestor to the $z-1$ ancestors of each
$n$-tree (see figure \ref{fig1}).
In order to get a full tree, one has to
``glue'' $z$ half-space-trees instead of
$z-1$ at the last step. Let $P_n^{\sigma}(M)$ be the
conditional probability
for a $n$-half-space-tree to have a
magnetization $M$, given the spin
of the ancestor is $\sigma$. Of course,
\begin{equation}
\sum_M P_n^{\sigma}(M) = 1
\end{equation}
The recursion relation for $P_n^{\sigma}(M)$ is
\begin{eqnarray}
\nonumber
P_n^{\sigma}(M) &=& \sum_{M_1, ... , M_{z-1}}
\delta(M-(M_1 + ... + M_{z-1}+\sigma))\\
\label{eq1}
&&
\sum_{k=0}^{z-1} {{z-1} \choose k} x^{k} (1-x)^{z-k-1}
\prod_{i=1}^{k} P_{n-1}^{-\sigma}(M_i)
\prod_{i=k+1}^{z-1} P_{n-1}^{\sigma}(M_i)
\label{eq23}
,
\end{eqnarray}
where $x$ is the probability for breaking one bond:
\begin{equation}
x = \frac{e^{- \beta J}}{e^{\beta J}+ e^{- \beta J}}
{}.
\end{equation}
The initialization of the recursion is given by
$ P_1^{\sigma}(\sigma')=\delta_{\sigma,\sigma'}$.
This recursion can be performed numerically,
at least for a small number
of generations. The result is plotted on
figure \ref{fig2} for $z=3$ and
10 generations.
For a finite size tree and at low temperature,
the magnetization distribution
presents a non gaussian structure, reminiscent
of the magnetization
distribution of the central spin in Bethe lattice
spin glasses \cite{4}
\cite{8}. Notice that the temperature which
controls the departure
from the Gaussian distribution is lower than
the bulk transition temperature.

In order to determine precisely this temperature,
we compute the
recursion relations for the average magnetization.
Using the recursion relations (\ref{eq1}), we find
\begin{equation}
\langle M \rangle_{n+1}^{+} = 1 + \sum_{k=0}^{z-1}
{{z-1} \choose k} x^{k} (1-x)^{z-k-1} \left(
k \langle N \rangle_n^{-} + (z-k-1)
\langle M \rangle_n^{+} \right)
\label{eq2}
\end{equation}
It is clear that, since $P_1^{+}(\sigma)
= P_1^{-}(- \sigma)$,
 for all n $P_n^{+}(M) = P_n^{-}
(-M)$, so that
$\langle M \rangle_n^{+} + \langle M
\rangle_n^{-} = 0$.
Putting this equation into (\ref{eq2})
and using the well-known
relations for the sum of binomial series,
we obtain
$\langle M \rangle_{n+1}^{+} = p \langle
M \rangle_n^{+} + 1$ and
$\langle M \rangle_0^{+} = 1$
where $p$ is defined as $p = (z-1) \tanh{\beta J}$.
This recursion can be easily solved, and one gets
\begin{equation}
\frac{\langle M \rangle_n^{+}}{N_n} =
\frac{z-2}{p-1} \frac{p^{n+1}-1}{(z-1)^{n+1}-1}
,
\end{equation}
where $N_n$ is the number of sites of a
$n$-half-space-tree, given by
(\ref{eq22}).
The structure of the distribution of
magnetization is non-Gaussian
provided $2 x n \ll 1$,
that is
\begin{equation}
\label{eq16}
T < T_g=
\frac{J}{\ln{n}}=
\frac{J}{\ln{\left(\ln{N_n}/\ln(z-1)\right)}}
{}.
\end{equation}
It is clear that the temperature $T_g$
(which will be identified wit the glass
temperature) decreases very slowly with
the system size. For instance,
in limit where
$N \simeq 6.02.10^{23}$
and $z=3$, $T_g = J/4.4$.
We conclude from this analysis that
finite-size effects persist in
the limit of a macroscopic number of sites. For $T_g$ to be
drastically reduced,
one should consider systems of size
$\exp{6.02.10^{23}}$ (!).
Since the appearance of glassiness
is a finite-size effect,
we stress that $T_g$ is
a {\sl cross-over} temperature
scale even for macroscopic systems.

\subsection{Structure of the Magnetization
Distribution for $T < T_g$}
We would like to understand qualitatively the
structure of the maxima of
the magnetization probability distribution
below $T_g$; more specifically we want
localize the maxima and to calculate their weight.
To do this, we use the normalized continuous
magnetization variable
$m = M/N_n \in [-1,1]$ and the
associated density
$\rho_n(m) = N_n P_n^{+}(M)$.
The recursion relations for $\rho_n(m)$
are derived in a straightforward fashion
from those for $P_n^{+}(M)$ that are shown
in
(\ref{eq1}). Since this relation
is a convolution, we write the recursion
in terms of the Fourier transform
$\tilde{\rho}_n(k)$ of $\rho_n(m)$:
\begin{equation}
\tilde{\rho}_n(k) = \int_{- \infty}^{+ \infty}
e^{i k m} \rho_n(m) dm
\end{equation}
We obtain the recursion of the $\tilde{\rho}$'s:
\begin{eqnarray}
\tilde{\rho}_{n+1}(k) &=&
\left( x \tilde{\rho}_n(-\frac{k}{z-1})
+ (1-x) \tilde{\rho}(\frac{k}{z-1}) \right)^{z-1}\\
\tilde{\rho}_0(k) &=& e^{i k}
\end{eqnarray}
We now study the special case $z=3$
where the formulae are simpler.
It is easy to check by recursion that
\begin{equation}
\label{eq5}
\tilde{\rho}_n(k) = (1 - 2 x (2^{n}-1))
e^{i k}
+ 2 x \sum_{\alpha=0}^{n-1} 2^{\alpha}
\exp{\left(i(1-\frac{1}{2^{\alpha}}) k \right)}
+ O(x^{2})
{}.
\end{equation}
As we shall see later, this low-temperature
expansion
is meaningful below $T_g$ even in the
presence of a finite density
of kinks.
The expansion (\ref{eq5}) tells us that
$\rho_n(m)$ has peaks for
$m_{\alpha} = 1 - 1/2^{\alpha}$,
where $\alpha \in \langle 0, ..., n-1 \rangle$.
Moreover, we find that
the weight of the peak $\alpha+1$ is twice that of $\alpha$.
Inspection of figure \ref{fig2} indicates that
this prediction is correct,
at least in the region where the overlap between
the peaks is small.

These results can also be interpreted in the following fashion.
The expansion (\ref{eq5})
at order $x$ means that the magnetization density
is calculated
at the order of one kink. It is clear that a single
kink at generation
$n-\alpha$ leads to a magnetization $1 - 1/2^{\alpha}$
and that the number
of choices to put a kink at generation $n-\alpha-1$
is twice the number of
choices to put a kink at generation $n-\alpha$,
which is the content
of equation (\ref{eq5}). What is striking is
that below $T_g$, this
one kink picture is valid, even though we deal
with a finite density of
kinks $x$. This means that $T_g$ is the
temperature below which the
kinks are rarely nested. In order to check
this assertion, we determine
the condition for when the kinks to "induce''
well-defined domains of
flipped spins; this criterion will determine
the validity of
the expansion (\ref{eq5}). The overlap between
the domains induced
by the kinks is small provided
$x N_n \langle S \rangle_n < N_n$.
In this expression, $\langle S \rangle_n$ is
the average size of
a domain of flipped spins induced by a single
kink. The number of
descendants of a kink at level $n-p$ for a
$n$-half-space-tree
is $S_{p,n} = 1+2+...+2^{n-p}=2^{n-p+1}-1$.
The average over $p$ of $S_{p,n}$ is
\begin{equation}
\langle S_{p,n} \rangle_n =
\frac{\sum_{p=1}^{n} 2^{p} S_{p,n}}
{\sum_{p=1}^{n} 2^{p}} = n \frac{2^{n}}
{2^{n}-1} -1 \simeq n-1
{}.
\end{equation}
The condition for ``non-overlap'' is  $x n < 1$
which is just $T < T_g$. We conclude that for
$T< T_g$, the number of kinks is
small enough for the system to develop
well-defined domains of flipped spins.
In this temperature regime,
the excitations of the spin
system are kinks which
are local in the bond variables  but  are {\sl highly}
non-local in terms of the spins.
in terms of bounds, but highly non-local
in terms of spins.

\subsection{Magnetization Distribution
in a Magnetic Field}
The magnetization distribution in a
magnetic field can be computed
using
\begin{equation}
P^{\sigma}_{n,h}(m) =
\frac{P^{\sigma}_{n,0}(m) e^{\beta m h}}
{\sum_{m'} P^{\sigma}_{n,0}(m')e^{\beta h m'}}
{}.
\end{equation}
The conditional magnetization distribution
(the central spin being
parallel to the field) is found to converge
rapidly to a gaussian
as the field increases. Since large domains
of flipped spins
do not survive in a magnetic field, the
cross-over field is
expected to decrease drastically with the
system size. Indeed, we have
shown previously that the cross-over field
$h_{co}$ of equation
(\ref{eq24}) decreases exponentially with
the number of generations.
The distribution of magnetizations is plotted on
figure \ref{fig23}, where the central spin
is taken to be parallel to the  field.


\section{Barrier Structure}
We have shown that the excitations below
$T_g$ are broken bonds.
In order to characterize the dynamics,
we calculate
the barriers associated with these excitations.
The energy barriers at zero temperature
 of a half-space-tree are
defined as follows: one starts with a
configuration where all
the spins are up and then
one considers single-spin-flip paths
from the initial configuration
to a final one where all the spins
are reversed.
To each such single-spin-flip path,
we associate the maximal
energy reached during the "passage'' from
the initial to the final configuration
where we take the energy of the
former to be zero.
Then the barrier is defined
as the minimum over all the paths of the
maximum energy of one path.
Typically, the Monte Carlo algorithm
samples all the paths in an ergodic
way; by contrast the Swenson algorithm does not generate
paths with respect to the single spin flip.
In Appendix B we give details of the calculation
of the barriers associated with the
$n$-half-space-tree.
Here we calculate the number of states
with a given barrier $E^{b}_{\alpha}$ at a given
temperature below $T_g$, for a full $n$-tree.
Following our previously established convention,
the leaves and the center reside at the first
and the $n$-th generation respectively.
We note there exist $n^{*}$
kinks at a given temperature
where
$n^{*} = N x$.
The barrier for a configuration of $n^{*}$
kinks is assumed to be
only a function of the
generation $\alpha$ of the kink which is the
closest to the origin.
In order to calculate the number of states
with an energy barrier
$E_{\alpha}^{b}$, we have to enumerate all
the configurations
with no kink between generation $\alpha + 1$
and $n$, $n_{\alpha}$ kinks
at generation $\alpha$ and $n^{*}-n_{\alpha}$
kinks between generations
$\alpha-1$ and $1$. We call
the number of such configurations
of kinks $g(\alpha)$. We make the approximation that the
energy barrier of all these
configurations is $E_{\alpha}^{b}$ so that
its life-time is,
according the the Arrhenius law,
\begin{equation}
\tau_{\alpha} = \tau_0 \exp{\left(\lambda
\beta E_{\alpha}^{b}\right)}
,
\end{equation}
where $\lambda$ is a constant.
Such a configuration of kinks is displayed
on figure \ref{fig3}.
Clearly, we have
\begin{equation}
\label{eq8}
g(\alpha) = \sum_{n_{\alpha}=1}^{n^{*}}
g_{\alpha}(n_{\alpha})
,
\end{equation}
with
\begin{equation}
g_{\alpha}(n_{\alpha}) = {z(z-1)^{n-\alpha-1}
\choose n_{\alpha}}
{\frac{z}{z-2} \left( (z-1)^{n-1}-
 (z-1)^{n-\alpha} \right)
\choose n^{*}-n_{\alpha}}
{}.
\end{equation}
We can calculate the sum and obtain
\begin{equation}
g(\alpha) = {\frac{z}{z-2}\left(
(z-1)^{n-1}-(z-1)^{n-\alpha-1}\right)
\choose n^{*}}
-
{\frac{z}{z-2} \left( (z-1)^{n-1}-
(z-1)^{n-\alpha} \right) \choose n^{*}}
,
\end{equation}
and we get the probability $P(\alpha)$ for
the system to be in a valley
with a barrier $E_{\alpha}$:
\begin{equation}
P(\alpha) = {N \choose N x}^{-1} \left[
{N(1-(z-1)^{-\alpha}) \choose Nx}-
{N(1-(z-1)^{-\alpha+1})\choose Nx} \right]
,
\end{equation}
where we have normalized by the total
number of accessible states
at a given temperature on a $n-$tree.
We have also approximated the number
of sites as follows:
\begin{equation}
N = \frac{z(z-1)^{n-1}-2}{z-2} \simeq
\frac{z(z-1)^{n-1}}{z-2}
{}.
\end{equation}
Assuming both that $x \ll 1$, and that
$(z-1)^{- \alpha} \ll 1$, and using
Stirling's formula, we get
\begin{equation}
P(\alpha) \simeq \exp{\left( -
\frac{N x}{(z-1)^{\alpha}} \right) }
- \exp{\left( - \frac{N x}
{(z-1)^{\alpha-1}} \right)}
{}.
\end{equation}
If $N z /(z-1)^{\alpha} \ll 1$, we obtain
\begin{equation}
\label{eq11}
P(\alpha) \simeq N x \frac{z-2}
{(z-1)^{\alpha}}
{}.
\end{equation}
As expected, $P(\alpha)$ decreases
 as a function of $\alpha$, which indicates
that the long-lived states are less
numerous than their short-lived counterparts.

\section{Glauber Dynamics}
We now study the Glauber dynamics of
the spin system on the Cayley tree.
\subsection{The Glauber Matrix}
We begin with a general discussion of
Glauber dynamics.
Let $p(\{\sigma\},t)$ be the probability
for the system to be in a state
$\{\sigma\}$ at time $t$. Following
Glauber \cite{ref3}, we define
$w_i(\{\sigma\})$ as the probability
per unit time that the spin $i$
flips from $\sigma_i$ to $-\sigma_i$,
 while the others remain fixed.
The master equation is
\begin{equation}
\label{eq10}
\frac{d}{dt} p(\{\sigma\},t) = -
\left( \sum_{i=1}^{N}
w_i(\{\sigma\}) \right) p(\{\sigma\},t)
+ \sum_{i=1}^{N}
w_i(\{\sigma_1,...,-\sigma_i,...,\sigma_N\})
p(\{\sigma_1,...,-\sigma_i,...,\sigma_N\},t)
{}.
\end{equation}
Since we require the Boltzman distribution
 to be a fixed point, the
coefficients $w_i(\{\sigma\})$ are of the form
\begin{equation}
w_i(\{\sigma\}) = \frac{1}{2}(1-\sigma_i
\tanh{(\beta J
\sum_{j \in V(i)} \sigma_j)})
,
\end{equation}
where $V(i)$ is the set of neighbors of
the site $i$.
If one denotes by ${\bf p}(t)$ the $2^{N}$
vector of the $p(\{\sigma\},t)$,
equation (\ref{eq10}) can be written as
\begin{equation}
\frac{d}{dt} {\bf p}(t) = {\bf G}.{\bf p}(t)
\label{eq25}
,
\end{equation}
where ${\bf G}$ is the Glauber matrix.
We first show some properties of the matrix
${\bf G}$.
Since the Boltzman distribution is a steady
state of the dynamics, its
corresponding eigenvalue is zero whatever
the temperature.
Even though it is {\sl not} symmetric,
the matrix {\bf G} can be diagonalized and
its eigenvalues are
real; we give the proof of this statement
here.
The Glauber
matrix satisfies detailed balance
which means that
${\bf G}_{\alpha,\beta}
{\bf p}^{(0)}_{\beta} =
{\bf G}_{\beta,\alpha}
{\bf p}^{(0)}_{\alpha}$ where ${\bf p}^{(0)}$ is
the Boltzman distribution.
As a consequence
\begin{equation}
\left( {\bf p}_{\alpha}^{(0)} \right)^{1/2}
 {\bf G}_{
\alpha \beta} \left(
{\bf p}_{\beta}^{(0)} \right)^{1/2}
=\left( {\bf p}_{\beta}^{(0)} \right)^{1/2}
 {\bf G}_{
\beta \alpha} \left( {\bf p}_{\alpha}^{(0)}
\right)^{1/2}
{}.
\end{equation}
Let us define a matrix ${\bf M}$ such that
\begin{equation}
{\bf M}_{\alpha \beta} = \left(
{\bf p}_{\alpha}^{(0)} \right)^{1/2}
{\bf G}_{\alpha \beta}
\left({\bf p}_{\beta}^{(0)} \right)^{1/2}
{}.
\end{equation}
Then  ${\bf M}$ is symmetric. Let ${\bf p}$
be a right eigenvector of
the Glauber matrix. Then
\begin{equation}
\sum_{\beta} {\bf G}_{\alpha \beta}
{\bf p}_{\beta} = \lambda {\bf p}_{\beta}
\end{equation}
is equivalent to
\begin{equation}
\sum_{\beta} {\bf M}_{\alpha \beta}
\left( {\bf p}_{\beta}^{(0)}
\right)^{-1/2} {\bf p}_{\beta} =
 \lambda \left(
{\bf p}^{(0)}_{\alpha} \right)^{-1/2} {\bf p}_{\alpha}
,
\end{equation}
so that $\left({\bf p}_{\alpha}^{(0)}
 \right)^{-1/2}
{\bf p}_{\alpha}$ is an eigenvector of
${\bf M}$. We conclude that ${\bf G}$
is diagonalizable, and that all of its
eigenvalues are real.

The spectrum in the infinite temperature
limit can be understood in the
following manner. If we define a state
vector $| \psi \rangle$ by
\begin{equation}
| \psi \rangle = \sum_{\{\sigma\}}
f((\{\sigma\})
|\sigma_1 \rangle \otimes ... \otimes
|\sigma_N \rangle
,
\end{equation}
then its  dynamics are
\begin{equation}
\frac{d}{dt} |\psi \rangle = -
\frac{N}{2} |\psi \rangle
+ \frac{1}{2} \sum_{i=1}^{N}
\sigma_i^{x}|\psi \rangle
,
\end{equation}
so that the eigenvalues of the
Glauber matrix at infinite temperature
are of the form
\begin{equation}
\lambda = -\frac{N}{2} +
\frac{1}{2} \sum_{i=1}^{N} \mu_i
,
\end{equation}
where $\mu_i = \pm 1$.
The spectrum in the infinite
temperature limit is composed
of levels at integer values
between $-N$ and $0$, with a
degeneracy given by the binomial
coefficients.

For bipartite lattices,
such as the square lattice of the Caley
tree, the spectrum of the matrix ${\bf G}$
is symmetric; more specifically
if $\lambda$ belongs to the
spectrum, then $-N-\lambda$
is an eigenvalue too. We give the
proof of this statement now.
 Let ${\bf X}\{\sigma\}$
be an eigenvector of ${\bf M}$, with an
eigenvalue $\lambda$:
\begin{equation}
\lambda {\bf X}\{ \sigma \} =
- \sum_{i=1}^{N} \frac{1}{2}(1 -
\sigma_i \tanh{(\beta J h_i)}
{\bf X}\{\sigma\} + \sum_{i=1}^{N}
\frac{1}{2 \cosh{\beta J h_i}}
{\bf X}\{\sigma_1, ..., -
\sigma_i,...,\sigma_N\}
,
\end{equation}
where $h_i$ is defined by
\begin{equation}
h_i = \sum_{j \in V(i)} \sigma_j
{}.
\end{equation}
Let ${\bf Y}\{ \sigma \}$ be defined as
\begin{equation}
{\bf Y}\{\sigma\} = (-1)^{\nu\{\sigma\}}
{\bf X}\{\tilde{\sigma}\}
,
\end{equation}
where $\nu\{\sigma\}$ is the number of up
spins in the configuration
$\{\sigma\}$.
$\{\tilde{\sigma}\}$ is deduced from
$\{\sigma\}$ by flipping the spins
of one of the two sublattices. Then,
\begin{eqnarray}
\nonumber
({\bf M}{\bf Y})\{\sigma\} &=&
- \sum_{i=1}^{N} \frac{1}{2}\left( 1-\sigma_i
\tanh{(\beta H h_i)} \right)
(-1)^{\nu\{\sigma\}} {\bf X}\{\tilde{\sigma}\}
+ \sum_{i=1}^{N} \frac{ (-1)^{ \nu
\{ \sigma_1,...,-\sigma_i,...,\sigma_N \} }}
{2 \cosh{ (\beta J h_i) }} {\bf X}
\{\tilde{\sigma}_1,...,
-\tilde{\sigma}_i,...,\tilde{\sigma}_M\}\\
&=& (-1)^{\nu\{\sigma\}} \left[
- \sum_{i=1}^{N} \frac{1}{2}
\left(1+\tilde{\sigma}_i
\tanh{(\beta J h_i)} \right)
{\bf X}\{\tilde{\sigma}\}
- \sum_{i=1}^{N} \frac{1}{2 \cosh{(\beta J h_i)}}
{\bf X}\{\tilde{\sigma}_1,...,-
\tilde{\sigma}_i,...,\tilde{\sigma}_N\}
\right]\\
&=&-(N+\lambda) (-1)^{\nu\{\sigma\}}
{\bf X}\{\tilde{\sigma}\}
= -(N+\lambda) {\bf Y}\{\sigma\}
{}.
\end{eqnarray}
Given an eigenvector ${\bf X}$ for the
eigenvalue $\lambda$,
we have constructedd an eigenvector ${\bf Y}$
for the eigenvalue $-N-\lambda$.

The difference between (\ref{eq25}) and
the Schr\"odinger equation is that
quantum mechanics preserves the scalar
product which  results  in Hermitian
Hamiltonians. Furthermore,
the physical states ``reside'' in a Hilbert space,
and each state of this Hilbert space is
physical. In the case of the Glauber
matrix there is no such vectorial space;
more specifically, the sum of
two probability distributions is not a
probability distribution.
However some quantities are
conserved by the dynamics. It is easy
to show that the eigenvectors
of ${\bf G}$ for the non-zero eigenvalues
have the property that
\begin{equation}
\sum_{\{\sigma\}} {\bf p}\{\sigma\} = 0
{}.
\end{equation}
This is a simple consequence of the fact that
the Glauber matrix preserves the quantity
\begin{equation}
\sum_{\{\sigma\}} {\bf p}\{\sigma\}
{}.
\end{equation}

\subsection{Spectrum of the Glauber Matrix}
We now discuss the spectrum of
the Glauber dynamics as a function
of temperature. We have performed
a series of numerical
diagonalizations on small clusters,
and then speculate on the behavior for larger
 systems specifically with respect to the statistics
of their eigenvalues. Since this aspect of the
Glauber dynamics has not studied in much detail
before, we examine the Ising model on one- and two-
dimensional lattices with
this approach before
ending with the Cayley tree;
in particular, we examine their relaxation spectra
as a function of temperature and
attempt to extrapolate their behavior in
the thermodynamic limit.

We begin with the Ising chain,
where we have determined the spectrum
as a function of temperature for
a six-site system with
open boundary conditions;
the result is displayed in  figure
\ref{fig4}. We also treated the case of a
10 sites ring. Of course,
we did not follow the 1024 levels as a
function of temperature,
but computed the eigenvalues for a given
temperature.
The spectrum is plotted on figure \ref{fig5}.
 From these results,
we can speculate on its behavior for
large systems. The spectrum
is {\it completly integrable} in the sense
that no level repulsion occurs
in the evolution of the eigenvalues as a
function of temperature.
It is clear from figure \ref{fig4} that
the evolution of $1/\tau_i$
is monotonous as a function of $\beta$
and that no repulsion is observed.
However, the distribution of eigenvalues
is non uniform, since
a curvature can be seen on figure \ref{fig5}.
It is not surprising that the
spectrum reveals an underlying integrable
 dynamics
since the dynamics was indeed
integrated by Glauber in \cite{ref3}.
We also note the existence of a gap at non-zero
temperature, which vanishes
in the limit of zero temperature;
here the
eigenvalue is doubly degenerate,
corresponding to the existence
of a symmetry-breaking phase at $T=0$.

We have also
studied small clusters (of size 3x3)
of the two-dimensional Ising model;
the resulting spectrum in the low-temperature phase
is plotted on figure \ref{fig6}.
The obvious difference with the results for
the one-dimensional chain is the
presence of clusters of relaxation
times in the low-temperature spectrum.
The evolution of the eigenvalues as a
function of the inverse temperature
is plotted for a small 3x2 cluster on
figure \ref{fig7}.
Level repulsion is visible in the
vicinity of the transition temperature.
For an infinite size system,
one expects the presence of a gap between
zero  and the first negative eigenvalue
in the spectrum  for $T > T_c$;
it should vanish at the
transition temperature, leading to
a doubly-degenerate zero eigenvalue
in the interval $[0,Tc]$ which corresponds
to the existence of a symmetry-breaking
stable phase.
Thinking of the spectrum as a function of $\beta$,
one expects the spectrum to
display chaotic behavior in the vicinity of the
transition temperature, leading
to a rearrangement of the inverse
relaxation times
at low temperature and the presence
of clusters of relaxation times in the
zero-temperature limit.
However, it is not clear whether the
level-spacing statistics should be
Gaussian Orthogonal Ensemble (G.O.E.) in character
 in an appropriate sector;
one might be left with a mixture
of G.O.E. and Poisson statistics.
A better understanding of these
level-spacing
statistics would involve the search of
the symetry sectors using group
theory which will be pursued in a future project.

In our study of the Cayley tree, we
determined the spectrum of eigenvalues
as a function of temperature
for $z=3$ nad $n=1$,
which corresponds to four  sites. The
resulting ``spaghetti''  are plotted in figure
\ref{fig8}. Though the system
size is
very small, one can
still see repulsion of the eigenvalues
in the vicinity of the bulk transition
temperature. We determined the spectrum
of the Glauber matrix
at low temperatures for
$n=2$, resulting in 1024 eigenmodes;
the eigenvalues are
plotted on figure \ref{fig9}.
The spectrum displays
clusters but, unlike the case of the
two-dimensional Ising model,
these clusters are equidistant and
occur  at integer values.
Furthermore  the $\lambda=0$
cluster contains 8 eigenvalues whereas
it contained just two
for the 2D Ising model.
For the case of $N=10$ and $\beta=2$,
shown in figure \ref{fig9},
we find that $n^{*} \simeq 0.18 < 1$.
In the previous section, we conclude from equation(\ref{eq11})
that $P(\alpha)$ is independent of the temperature;
this result is confirmed
by the spectrum of the Glauber matrix
since the $\lambda=0$ and the
$\lambda = -1$ clusters
are separated by a gap.

\subsection{Determination of the Glauber Dynamics}
We now diagonalize the Glauber matrix in order
to determine the dynamics of the model in an explicit
fashion.
Let ${\bf u}_{\alpha}$ be the eigenvectors
of ${\bf G}$:
${\bf G}.{\bf u}_{\alpha} = \lambda_{\alpha}
{\bf u}_{\alpha}$
,
and let ${\bf P}$ be the passage matrix from
the natural basis of
pure states ${\bf e}_{\alpha}$ to the basis
${\bf u}_{\alpha}$:
\begin{equation}
{\bf u}_{\alpha} = \sum_{\beta}
{\bf P}_{\alpha \beta}
{\bf e}_{\beta}
{}.
\end{equation}
We look for the temporal evolution of
the states ${\bf e}_{\alpha}$.
At the initial time (t=0)
\begin{equation}
{\bf p}_{\alpha}(0)={\bf e}_{\alpha} =
\sum_{\beta} {\bf P}_{\alpha \beta}^{-1}
{\bf u}_{\beta}
{}.
\end{equation}
and  at $t>0$, the state is a mixture
of pure states and is given by
\begin{equation}
{\bf p}_{\alpha}(t) = \sum_{\beta}
 {\bf P}_{\alpha \beta}^{-1}
e^{\lambda_{\beta} t} \sum_{\gamma}
{\bf P}_{\beta \gamma}
{\bf e}_{\gamma}
{}.
\end{equation}
We can easily compute the magnetization
of ${\bf p}_{\alpha}(t)$.
We applied this procedure to the case of a
tree with one generation.
The evolution of the magnetization of the
16 pure states is plotted on figure
\ref{fig10} at low temperature. Of course,
on very long time scales,
the magnetization of all the states relaxes
to zero, due to the fact that
the eigenvalues associated to the symmetry
breaking state are not strictly
zero. The evolution of the pure states
indicate the existence of metastable
states, which are the precursors of the
metastable states present
for larger values of the number of generations.

\subsection{The Factorization Approximation}
We recover the bulk critical temperature
by considering the asymptotics
of a simplified Glauber dynamics.
Glauber has shown \cite{ref3} that one can
replace the $2^{N}$ variables of
the linear dynamics by a hierarchy of $N$,
nonlinear, coupled equations
for the correlations functions. This
 procedure is very similar to the
transformation of the Liouville equation
into the B.B.G.K.Y. hierarchy
in the kinetic theory.
The first equation of the hierarchy is
\begin{equation}
\frac{d}{dt} q_i(t) = - q_i(t) +
\langle \tanh{( \beta J
\sum_{j \in V(i)} \sigma_j )} \rangle
{}.
\end{equation}
In this expression, $q=\langle \sigma \rangle$
and $V(i)$ is the set
of neighbors of site $i$.
In the case of a one dimensional chain,
one can use the fact that
\begin{equation}
\langle \tanh{\beta J(\sigma_{i-1}+
\sigma_{i+1})} \rangle=
\frac{1}{2} \tanh{( \beta J)}
(q_{i-1}+q_{i+1})
,
\end{equation}
and one gets a closed equation for the
1-point correlation functions.
It is also clear that the whole hierarchy decouples,
 and that one
can use this decoupling to integrate the dynamics.
In the case of a
$z=3$ tree, one has to take into account the fact
that the sites inside
the tree have three neighbors;by contrast the leaves
 have one
neighbor. For this coordination,
\begin{equation}
\tanh{\beta J(\sigma_1+\sigma_2+\sigma_3)} = \alpha
(\sigma_1+\sigma_2+\sigma_3) + \gamma (\sigma_1 + \sigma_2
+ \sigma_3)^{3}
,
\end{equation}
where the coefficients $\alpha$ and $\gamma$
 are determined by
\begin{eqnarray}
\alpha &=& \frac{1}{24} (27 \tanh{\beta J} -
\tanh{3 \beta J})\\
\gamma &=& \frac{1}{24} (\tanh{3 \beta J} -
 3 \tanh{\beta J})
{}.
\end{eqnarray}
We can thus obtain the first equation of the
hierarchy in the case of the
$z=3$ tree. For the sites with three neighbours,
\begin{equation}
\frac{d}{dt} q_i = - q_i + (\alpha+7 \gamma)
 \sum_{j \in V(i)} q_j
+ 6 \gamma \langle \prod_{j \in V(i)} \sigma_j \rangle
\end{equation}
For the leaves of the tree
\begin{equation}
\frac{d}{dt} q_i = - q_i + q_j \tanh{\beta J}
,
\end{equation}
where $j$ is the neighbour of $i$.
The factorization approximation consists in
 decoupling the third order
correlations into
\begin{equation}
\langle \prod_{j \in V(i)} \sigma_j \rangle
= \prod_{j \in V(i)} q_j
{}.
\end{equation}
This approximation leads to the bulk behavior
 in the high temperature
phase and in the vicinity of the transition.
We start from a configuration of spins such as
 $q_i(0)=q_j(0)$ if
the sites $i$ and $j$ belong to the same generation.
Then, for $t>0$, $q_i(t)=q_j(t)$ if we work
with the factorized
dynamics. The factorized dynamics depends
 only on $n$
variables, one per generation, and is
\begin{eqnarray}
\nonumber
\frac{d q_n}{dt} &=& - q_n + 3(\alpha +
7 \gamma) q_{n-1} +
6 \gamma q_{n-1}^{3}\\
\label{eq12}
\frac{d q_i}{dt} &=& - q_i + (\alpha +
7 \gamma) (2 q_{i-1} + q_{i+1})
+ 6 \gamma q_{i-1}^{2} q_{i+1}\\
\nonumber
\frac{d q_1}{dt} &=& - q_1 + q_{2}
\tanh{\beta J}
,
\end{eqnarray}
where $i$ runs from 2 to $n-1$.
The equilibrium properties are calculated
from the asymptotic values of the
dynamics. We note that the factorized
dynamics possesses a non-trivial
fixed point for a finite size system,
whereas the complete dynamics
possesses only the Boltzman distribution
as a fixed point. The
asymptotics of the factorized dynamics
is found to reproduce quite
well the bulk properties of the tree.
On figure \ref{fig11}, $q_n(+ \infty)$
is plotted as a function of the inverse temperature.
This curve is in agreement
with the fact that the central spin exhibits a
 mean field like transition
at $\beta_c \simeq 0.54$.
We also plotted the asymptotic magnetization
 of the $m$ generations
which are the closest to the central spin.
It is clear that the predictions of the
factorized dynamics are qualitatively
wrong as soon as one goes out from the center.
For instance,
it is clear that the entire tree does not
develop a transition
at $\beta = \beta_c$.
We note that, in equations (\ref{eq12}), the
transition temperature is determined
entirely by the stability of
the zero fixed point of the linear problem.
Below $T_c$, the largest eigenvalue
of the linear problem is positive, and
negative above $T_c$.
As in the Ginzburg-Landau
 theory,
the nonlinear terms are responsible for
the maximum bound on the dynamical variables
in the low- temperature phase.

\section{Monte Carlo Dynamics}
\subsection{Relaxation of a Single kKink}
We begin by considering
the case of a single kink and
look for the relaxation
of this excitation at low temperatures.
Let's call $K$ the set of
descendants of the kink. At time $t=0$,
the configuration
of kinks is such as $\sigma_i = -1$ is
$i \in K$ and $\sigma_i = 1$ if
$i \not \in K$. We follow the magnetization
of the spins of $K$ as
a function of time, for various size of $K$.
The result is plotted
on figure \ref{fig12}. The rapid relaxation
at small times
is attributed to the fact that the initial
state is not thermalized;
the thermalization occurs at small time
scales compared to the
collective processes of crossing the barrier.
We define the typical relaxation time $\tau$
as the time-scale associated with the vanishing of the magnetization.
In  Figure \ref{fig13} we have plotted
the logarithm of this relaxation time,$\ln{\tau}$,
as a function of the
number of generations of $K$ ($n_K$). The points are
approximately aligned,
indicating that the Arrhenius law
\begin{equation}
\tau \sim \tau_0 \exp{\left( \beta \lambda \frac{z-2}{2} n\right)}
\label{eq13}
\end{equation}
is well-satisfied.
The case $\lambda=1$ corresponds to the
relaxation of a $n$-half-space tree where
the ancestor is free. However, in our
 case, the ancestor is {\sl not} free
since the domain $K$ is connected to the
 remaining up spins. The barrier height
is essentially the same as in the case
$\lambda=1$, but the number of
paths to reverse the magnetization is changed.

\subsection{The A.C. Susceptibility}

Experimentally, the a.c. nonlinear susceptibility
has been a very useful probe of glassiness,\cite{ref4}
and we can apply the same techniques numerically to study
the tree problem that we are discussing here.
In parallel with what is done experimentally,
a small  a.c. external field $H(t)=h \sin{\omega t}$
is applied to the spin system,
and we measure
the magnetization response,
expanded into its
Fourier components
\begin{equation}
M(\omega,t) = \sum_{k \ge 0} \theta_k'
\sin{k \omega t}
+ \theta_k'' \cos{k \omega t}
,
\end{equation}
which has only odd harmonics. In practice,
we let the system relax
for four periods and we take measurements
only during the fifth one.
The in-phase susceptibilities are related
to the in-phase Fourier
coefficients as \cite{ref4}
\begin{eqnarray}
\theta_1' &=& \chi_1' h + \frac{3}{4}
\chi_3' h^{3} + \frac{5}{8} \chi_5' h^{5}
+ \frac{35}{64} \chi_7'h^{7} + ...\\
\theta_3' &=& \frac{1}{4} \chi_3' h^{3}
+ \frac{5}{16} \chi_5' h^{5}
+ \frac{21}{64} \chi_7' h^{7} + ...\\
\theta_5' &=& \frac{1}{16} \chi_5' h^{5}
+ \frac{7}{64} \chi_7' h^{7} + ...\\
\theta_7' &=& \frac{1}{64} \chi_7' h^{7}
+ ...
{}.
\end{eqnarray}
Since {\sl all} the nonlinear susceptibilities are divergent
at the critical temperature of a spin glass,\cite{ref4}
we must include all the measurable higher order harmonics
in order to properly include their influence on the lower
ones.\cite{ref4}.

The susceptibilities  $\chi_1'$  and $\chi_3'$
are plotted as a function of temperature
in Figures \ref{fig14} and \ref{fig15}
respectively for a number of frequencies.
Both $\chi_1'$ and $\chi_3'$ display
maxima, where that of the latter is more pronounced;
in a ``real'' spin glass in the thermodynamic limit
one would expect $\chi_1'$ to have a maximum,
and $\chi_3'$ to diverge.
We note that $\chi_3'$ is negative near its maximum,
which agrees with experiments
on spin glasses \cite{ref4}. The frequency
dependence of the position
of the maximum in $\chi_1'$ and $\chi_3'$
is also consistent with experiment,\cite{ref4} since
the temperature for which $\chi_1'$ and
$\chi_3'$ are maximum increases with
frequency. Thus this study
strongly suggests
that the Ising model on a finite Cayley tree
exhibits glassy-like behavior
at low temperatures;
we shall be more specific when we analyze
the Edwards-Anderson susceptibility
of this model.

\subsection{Autocorrelation Functions and Ageing}
A good test used to determine the presence of
glassiness in a given model involves
the computation of the autocorrelation functions \cite{ref5}
to determine whether they exhibit
ageing. The autocorrelation functions are
\begin{equation}
C(t,t_w) = \frac{1}{N} \sum_{i=1}^{N} \langle \sigma_i(t+t_w)
\sigma_i(t_w) \rangle - \langle \sigma_i(t+t_w) \rangle
\langle \sigma_i(t_w) \rangle
,
\end{equation}
where the sample is rapidly quenched below
$T_g$ from
a disordered high temperature state.
The Monte Carlo dynamics runs
from times $t = 0$ to $t = t_w$; the
autocorrelations are measured at $t=t_w$
where an average is taken over the initial
configurations.

The autocorrelations in the high-temperature
 phase are plotted on figure
\ref{fig17}. They decrease rapidly with the
time $t$ and are independent of
the waiting time, as expected;
this is simply a check of our code.
Figure \ref{fig18}
represents the autocorrelation functions
below $T_g$. The aucorrrelations {\it increase}
with increasing
$t_w$,
indicating  the presence of ageing.
Such behavior has been
observed in a wide class of glassy models,
including the
fully frustated hypercubic model \cite{ref5}.

\section{The Edwards-Anderson Order Parameter
and Susceptibility}

The recursive structure of the Cayley tree permits us to compute the
Edwards-Anderson order parameter and susceptibility
at all temperatures. This calculation has
already been done for the
$\pm J$ model with uncorrelated boundary
conditions in \cite{4}
\cite{8}. These authors found an Almeida-Thouless line for their
spin glass model, and
we wish to use the same technique to
analyse the model discussed here.
In order to compute the Edwards-Anderson order
parameter and susceptibility,
we consider one tree
plus one replica, with an intereplica
coupling $R$. If $\{ \sigma \}$
and $\{ \sigma' \}$ are the spin
configurations of the two replicas,
the Hamiltonian reads
\begin{equation}
H = - J \sum_{\langle i,j \rangle}
(\sigma_i \sigma_j + \tilde{\sigma}_i \tilde{\sigma}_j)
- R \sum_{i=1}^{N} \sigma_i \tilde{\sigma}_i
+ H \sum_{i=1}^{N} (\sigma_i + \tilde{\sigma}_i)
{}.
\end{equation}
As per usual with free boundary trees,
we have to distinguish between the properties of
the central spin and the whole tree;
in what follows we discuss both cases below.

\subsection{The whole tree}
The partition function can be calculated
 by the construction of figure
\ref{fig1}, even in the presence of the
intereplica coupling.
We take derivatives of the partition function with
respect to the intereplica coupling, $R$,
resulting in the Edwards-Anderson order
parameter
\begin{equation}
q_{EA} = \frac{1}{N} \langle \sum_{i=1}^{N}
 \sigma_i
\tilde{\sigma}_i \rangle = \frac{1}{N}
\frac{\partial}{(\partial
\beta R)} \ln{Z(R=0)}
\label{eq14}
\end{equation}
and
the Edwards-Anderson susceptibility:
\begin{equation}
\label{eq15}
\chi_{E.A.} = \frac{1}{N} \left(
\langle (\sum_{i=1}^{N} \sigma_i
\tilde{\sigma}_i)^{2} \rangle
- \langle \sum_{i=1}^{N} \sigma_i
\tilde{\sigma}_i \rangle^{2}
\right)
= \frac{1}{N} \frac{\partial}{\partial
(\beta R)^{2}} \ln{Z}(R=0)
{}.
\end{equation}
We now let $Z_{\sigma \sigma'}^{(n)}$
be the conditional partition function of
a $n$-half space tree
with respect to the ancestor's spins
$\sigma$ and $\sigma'$.
It is clear that
\begin{eqnarray}
 Z_{++}^{(n+1)} &=& e^{\beta R} e^{2 \beta H}
\left( e^{2 \beta J} Z_{++}^{(n)} + Z_{+-}^{(n)}
+ Z_{-+}^{(n)}
+ e^{- 2 \beta J} Z_{--}^{(n)} \right)^{z-1}\\
Z_{+-}^{(n+1)} &=& e^{-\beta R}
\left( Z_{++}^{(n)} + e^{2 \beta J} Z_{+-}^{(n)} +
e^{- 2 \beta J} Z_{-+}^{(n)}
+ Z_{--}^{(n)} \right)^{z-1}\\
Z_{-+}^{(n+1)} &=& e^{-\beta R}
\left( Z_{++}^{(n)} + e^{-2 \beta J} Z_{+-}^{(n)} +
e^{2 \beta J} Z_{-+}^{(n)}
+ Z_{--}^{(n)} \right)^{z-1}\\
Z_{--}^{(n+1)} &=& e^{\beta R} e^{2 \beta H}
\left( e^{-2 \beta J} Z_{++}^{(n)} + Z_{+-}^{(n)}
+ Z_{-+}^{(n)}
+ e^{2 \beta J} Z_{--}^{(n)} \right)^{z-1}
{}.
\end{eqnarray}
The initial conditions of the recursion are
$Z_{++}^{0} = e^{\beta R} e^{2 \beta H}$,
$Z_{+-}^{0} = Z_{-+}^{0} = e^{- \beta R}$ and
$Z_{--}^{0} = e^{\beta R} e^{- 2 \beta H}$.
For all $n$, we can show that $Z_{+-}^{(n)} =
 Z_{-+}^{(n)}$ and
we then define  $Z_1^{(n)} \equiv
Z_{++}^{(n)}$,
$Z_0^{(n)} \equiv Z_{+-}^{(n)} = Z_{-+}^{(n)}$ and
$Z_{-1}^n \equiv Z_{--}^{(n)}$.
Using this notation, the recursion
relations
 become
\begin{eqnarray}
Z_1^{(n+1)} = e^{\beta(R+2 H)}
 \left( e^{2 \beta J} Z_1^{(n)}
+2 Z_0^{(n)} + e^{- 2 \beta J}
Z_{-1}^{(n)} \right) ^{z-1}\\
Z_0^{(n+1)} = e^{- \beta R} \left( Z_1^{(n)}
+2 \cosh{\beta J} Z_0^{(n)} +
 Z_{-1}^{(n)} \right) ^{z-1}\\
Z_{-1}^{(n+1)} = e^{\beta(R-2 H)}
\left( e^{-2 \beta J} Z_1^{(n)}
+2 Z_0^{(n)} + e^{2 \beta J}
 Z_{-1}^{(n)} \right) ^{z-1}
{}.
\end{eqnarray}
which can be used
to obtain $q_{EA}$
and $\chi_{EA}$ in a straight-forward fashion . $q_{EA}$
and $\chi_{EA}$ are plotted in figures \ref{fig19}
and \ref{fig20} respectively for $n = 10$ and
$n = 80$.
In both cases, the curves corresponding to
these two system sizes are similar, even
though the number of sites
is small for $n=10$ (2047 sites) and
macroscopic for $n=80$ (4.02 moles
of sites).
We believe that this behavior is related to the very
slow variation of the
glass temperature $T_g$ with the system size.
In figure  \ref{fig20} we see clearly that
$\chi_{E.A.}$ does not diverge
in the thermodynamic limit but rather exhibits
a maximum, characteristic of a finite
size effect, even in the macroscopic regime.

\subsection{The central spin}
Following the authors of \cite{4}
\cite{8}, we note that
\begin{eqnarray}
Z^{(n)} &=& Z_1^{(n)} + 2 Z_0^{(n)} +
Z_{-1}^{(n)}\\
Q^{(n)} &=& \frac{1}{Z^{(n)}}
( Z_1^{(n)} - 2 Z_0^{(n)} + Z_{-1}^{(n)})
{}.
\end{eqnarray}
The Edwards-Anderson order parameter
of the central spin is
$\tilde{q}_{E.A.} = Q^{(n)}$,
and the Edwards-Anderson susceptibility is
\begin{equation}
\tilde{\chi}_{E.A.}^{(n)} =
\frac{\partial \tilde{q}_{E.A.}}
{(\partial \beta R)} (R=0)
\end{equation}
where the tilde symbol denotes quantities
with respect to the central spin.
The ``central spin''  Edwards-Anderson
susceptibility
$\tilde{\chi}_{E.A.}$ is plotted in figure
\ref{fig21} for various system sizes.
Though  we have seen that finite-size effects
are negligeable for the whole tree,
they become crucial for
the central spin.  More specifically,
the maximum values
of the Edwards
Anderson susceptibility of the central spin
does not increase (see figure \ref{fig22})
as a function of coordination $z$.
As the system size becomes
macroscopic, the glass temperature $T_g$
depends more on the coordination than
on the number of generation;
in particular, it increases with increasing
coordination consistent with equation (\ref{eq16}).

\subsection{Central spin in a magnetic field}
In the Sherrington-Kirkpatrick model, glassy
behavior persists even in the
presence of a magnetic field;
the Edwards-Anderson susceptibility
diverges with an exponent $\gamma=1$
through the Almeida-Thouless line.
We have studied the
behavior of the Ising n.n. ferromagnet (fm) on a Cayley tree
ferromagnetic model with free boundary
 conditions with an applied magnetic field.
Specifically we studied the variation of the
Edwards-Anderson susceptibility in a field as a function of system
size.
As displayed in figure \ref{fig25}, the maximum of
the Edwards-Anderson susceptibility
as a function of temperature depends on the size.
In figure \ref{fig26}, we showthe locus
of the maxima of
$\tilde{\chi}_{E.A.}$ for different magnetic
fields as a function of
temperature.
The Edwards-Anderson susceptibility
decreases strongly with increasing magnetic field,
indicating the vanishing of glassy behavior
consistent with expectation.

\section{Lee and Yang zeros}
The partition function of
the Ising fm on the Cayley tree in the absence of a magnetic
field is equal to that of the linear chain,
so that the Lee and Yang zeros in the
plane of $\exp{2\beta}$ have the identical structure
in both models.
Clearly this will not be the case for finite applied field.
We can introduce a complex magnetic field
and determine the
zeros of the partition function in
the complex plane of $\exp{2 \beta h}$.
Since we do not impose a uniform magnetic
field, but rather one only on the
leaves of the tree, we  can compute the
zeros using the
 recursion relations (\ref{eq17}),
which involve only second order
equations.
Since the magnetic field is non-uniform,
the underlying assumptions of the circle
theorem \cite{ref8} are not fulfilled.
 The zeros are not exactly on the
unit circle but areb in its  vicinity. More precisely,
numerical work indicates that the zeros
approach the unit circle with increasing generation. The zeros are plotted
on figure \ref{fig27} in the cases
$\beta < \beta_c$, $\beta = \beta_c$
and $\beta > \beta_c$. In figure
\ref{fig28}, we exhibit the ratio
 of the density of zeros in the
vicinity of $\exp{2 \beta h} = 1$ over the
averaged density on the circle.
The ratio is zero for temperatures larger
than the bulk transition
temperature, increases to a small positive value below the
bulk transition temperature,
and eventually becomes larger than unity below a
certain temperature.
In the case of a
Weiss model, the density of zeros $g(\theta)$
is given by \cite{ref9}
\begin{equation}
g(\theta) = \frac{1}{2 \pi} ( 1 - 2 r \cos{\varphi})
,
\end{equation}
where
\begin{eqnarray}
r &=& \sqrt{4 + r^{2} - 4 r \cos{\varphi}}
\exp{\left( - 2 \frac{T_c}{T}
 (1 - r \cos{\varphi})\right) }\\
\theta &=& - 2 \frac{T_c}{T} r
\sin{\varphi} + \varphi
+ \arctan{\left( \frac{r \sin{\varphi}}
{2 - r \sin{\varphi}}\right) }
{}.
\end{eqnarray}
The ratio of the density of
zeros at the point $\exp{2 \beta h}=1$ over
the average density of zeros can be determined numerically
from the sysetem of equations above, and is
plotted on the inset of figure (\ref{fig28}),
and is always less than
unity. By contrast on the tree, the density of zeros in
the vicinity of the
real axis is thus abnormally high below a
certain temperature,
which we identify with the temperature scale $T_g$.

\section{Discussion}

In summary, we have studied the static and the dynamical
properties of the nearest-neighbor Ising model on a
Cayley tree.  At zero field, we find that this
system displays glassy behavior below a size-dependent
temperature that scales inversely with the logarithm
of the number of generations; thus its glassy behavior
persists for a finite but macroscopic number of sites.  Because
the ratio of the number of surface to bulk sites, $\frac{N_s}{N}$, and
the strength of the external field, $h$,
play a key role in the physical behavior of the resident
spin system, the different thermodynamic limits associated
with the values of $\frac{N_s}{N}$ and $h$ are characterized;
the cross-over temperature, $T_g$, is associated with
fixed $\frac{N_s}{N}$ in the limit of vanishing applied
field.  Physically well-defined large domains of flipped
spins develop at $T_g$; at this temperature the probability
of nested spin clusters is small.  The largest
energy barriers associated with overturning these domains
is determined to scale logarithmically with the number
of sites at zero temperature, a result that should be
valid at finite, low temperatures if overlap between
spin clusters does not occur.  A dynamical study indicates
the appearance of metastable states and long relaxation
times at low temperatures.  The autocorrelations are computed after a waiting
time using
Monte Carlo dynamics; they exhibit ageing for $T < T_g$.
The temperature variations of the coefficients of $\chi'_1$
and $\chi'_3$ are also determined and they agree with
the existence of finite-size glassiness.  Finally the
Edwards-Anderson susceptibility of the entire tree
displays a maximum (but no divergence) that
evolves slowly with increasing system size; that of
the central spin has much more marked size-dependence.

We have thus performed a detailed characterization of
the low-temperature phase of a short-range periodic spin
model resident on a Cayley tree.  In this particular
case, we have found that it displays finite-size glassy
behavior that remains for a macroscopic number of sites;
perhaps it is best to characterize this low-temperature
phase as a very viscous spin liquid.  We note that
neither intrinsic disorder nor frustration exist due to
the initial Ising Hamiltonian;
the possibility of many low-temperature ``cluster''
states separated by very high energy barriers
is a direct consequence of the unusual topology
of the Cayley tree.  In many ways we hope that
this is a warm-up exercise towards the study
of spin models on more complicated non-Euclidean
lattices, e.g. on a constant triangulation associated
with a surface of negative curvature, where
the intrinsic geometry of the host may lead
to the possibility of glassiness in the absence of
both disorder and frustration.

\section{Appendix A}
We propose another derivation of the magnetization
 distribution for the
tree. Let $Z(\beta,h)$ the partition function
of the spin system in
the presence of an external field. Then,
\begin{equation}
P(M) = \frac{1}{Z(\beta,h)} \sum_{\{\sigma\}}
\delta \left( M - \sum_{i=1}^{N}\sigma_i \right)
\exp{\left(\beta ( J \sum_{\langle i,j \rangle}
\sigma_i \sigma_j
+ h \sum_{i=1}^{N} \sigma_i ) \right)}
{}.
\end{equation}
Using the Fourier representation of the delta function
\begin{equation}
\delta \left( M-\sum_{i=1}^{N} \sigma_i \right) =
\frac{1}{2 \pi}
\int_{0}^{2 \pi} d \lambda e^{i \lambda M}
e^{-i \lambda \sum_{i=1}^{N}
\sigma_i}
,
\end{equation}
we obtain
\begin{equation}
P(M) = \frac{1}{2 \pi} \int_{0}^{2 \pi}
d \lambda e^{i \lambda M} \frac{Z(\beta,h-i
\lambda / \beta)}
{Z(\beta,h)}
\label{eq9}
,
\end{equation}
where we have used the analytic continuation
of the partition
function for complex magnetic fields. This
 method is usefull provided
one knows how to calculte the partition
function, which is certainly feasible
on a tree.
We proceed by decimation, starting from the
border of the $n$-half-space-tree.
For future purpose, we note $Z_n(\beta,h,h_n)$
the partition function
of a $n$-half-space-tree with a magnetic field
$h$ acting on the spins of the
generations 0 to $n-1$, and $h_n$ on the spins
of the generation $n$. Then,
\begin{equation}
Z_n(\beta,h,h_n) = \left(4 (\cosh^{2}{(\beta J)}
+\sinh^{2}{(\beta h_n)}
\right)^{\frac{(z-1)^{n-1}}{2}}
Z_{n-1}(\beta,h,h+T.h_n)
,
\end{equation}
where the transformation of the magnetic field reads
\begin{equation}
T.h = \frac{z-1}{2 \beta} \ln{
\frac{\cosh{\beta(J+h)}}{\cosh{\beta(J-h)}}}
{}.
\end{equation}
The last term of the recursion corresponds
to the partition function of the
ancestor, which is simply
\begin{equation}
Z_1(\beta,h)= 2 \cosh{(\beta T^{n}.h)}
{}.
\end{equation}
It is straighforward to compute the partition
function using these
relations and to perform the Fourier transform
(\ref{eq9}) in order to
obtain the probability distribution of the magnetization.

\section{Appendix B}
In this appendix we give the value of the energy barrier $E(T_{n,z})$
for a half--space--tree $T_{n,z}$ with n generations and a coordination
number of $z$ for all sites except the root (coordination
$z-1$) and the leaves (coordination 1).
We also give the energy barrier $E(T^*_{n,z})$
for a complete tree $T^*_{n,z}$ with n generations and a coordination
number of $z$ for all sites except the leaves.
The derivation of the formula is due to A. Seb\H{o} and M. Preissmann,
and is published in extenso in \cite{aps}. Note that
the same problem arises in the VLSI circuit conception!
Generically the problem of finding the lowest energy
barrier is {\em NP-complete}, but the sub--problem of
finding the lowest energy barrier of a tree is polynomial,
and an explicit algorithm is given hereafter.
The value of the energy barriers are given by :
\begin{eqnarray}
\label {TTT}
c(T_{n,z}) = \lceil \frac{n(z-2)}{2} \rceil +1 \hspace{0.5 cm} (n,z \ge 3)\\
\nonumber
c(T^*_{n,z}) = \lceil \frac{(n-1)(z-2)}{2} \rceil +
		\lceil \frac{z-2}{2} \rceil
		+ 1
\hspace{0.5 cm} (n,z \ge 3)
\end{eqnarray}
In the above formula $\lceil x \rceil$ denotes the lowest of the integers
greater than $x$.
The demonstration of these formula is constructive. Firstly a
lower bound for $E$ is given. Then an algorithm is described which
produces a labeling of the sites. Flipping the spins in the
order of this labeling gives an energy barrier exactly equal to the
lower bound. The algorithm is recursive. It tries to produce
an optimal labeling of the sites where the root is labeled
{\em before} the configuration of highest energy is reached.
We call {\em strong} labeling such a labeling.
This extra constraint is useful when one applies $z$ times
the algorithm on a $T_{n,z}$ to compute $E(T_{n+1,z})$, or when
one applies the algorithm to $T_{n,z}$ and to $T_{n-1,z}$
to  compute $E(T^*_{n+1,z})$. A strong labeling does not
exist when $z$ and $n$ are both odd as shown in \cite{aps}.
Let us now consider on the case of $T_{n,z}$.
Formula \ref {TTT} means that
\begin{itemize}
\item when $z$ is even the increment in energy when
one goes from $T_{n,z}$ to $T_{n+1,z}$
is {\em constant} and equal to $\frac{z}{2}-1$
\item when $z$ is odd the increment in energy when
one goes from $T_{n,z}$ to $T_{n+1,z}$ is {\em alternatively}
$\frac{z-1}{2}-1$ and $\frac{z-1}{2}$
\end{itemize}

We give now the algorithm in the case of even $z$. The case
of odd $z$ is a slightly more complicated, but in the same spirit.
Consider $T_{n,z}$ as being made of $z-1$ copies of $T_{n-1,z}$
all of them connected to the site $0$. Each spin is identified
by two numbers $k,i$ with $0 \leq k < z-1$ and
$0 \leq i < N_{n-1}$ ($N_n$ is the number of sites
of $T_{n,z}$).
Let us note $\pi$ a strong labeling of $T_{n-1,z}$,
and $n_0$ the root of the $(\frac{z-2}{2}-1)$th copy of $T_{n-1,z}$.
The following labeling is a strong labeling of $T_{n,z}$:
\begin{enumerate}
\item \label {stp1} $(0,\pi(0)), \ (0,\pi(1)), \ldots, \ (0,\pi(N_{n-1}))$
\item \label {stp2} $\ldots$
\item $(\frac{z-2}{2}-1,\pi(N_n)), \ (\frac{z-2}{2}-1,\pi(N_n-1)),
	\ldots, \ (\frac{z-2}{2}-1,\pi(n_0))$
\item $0$
\item $(\frac{z-2}{2}-1,\pi(n_0-1)), \ (\frac{z-2}{2}-1,\pi(n_0-2)),
	\ldots, \ (\frac{z-2}{2}-1,\pi(0))$
\item $(\frac{z-2}{2},\pi(0)), \ (\frac{z-2}{2},\pi(1)),
	\ldots, \ (\frac{z-2}{2},\pi(N_{n-1}))$
\item \label {stp3} $\ldots$
\item \label {stp4}$(z-2,\pi(0)), \ (z-2,\pi(1)), \cdots, \
	(z-2,\pi(N_{n-1}))$
\end{enumerate}
Note that in step \ref {stp1}, \ref {stp2}, \ref {stp3}, and \ref {stp4}
any admissible permutation can be used instead of the strong labeling $\pi$.
It is shown in \cite{aps} that the above labeling
is indeed an optimal labeling
and it can be used to implement a recursive algorithm
to find a path between the two ferromagnetic states.



\newpage
\renewcommand\textfraction{0}
\renewcommand
\floatpagefraction{0}
\noindent {\bf Figure captions}

\begin{figure}[h]
\caption{}
\label{fig1a}
(a) A finite Cayley tree  and (b) a section  of a Bethe lattice with
coordination $z=3$.
\end{figure}

\begin{figure}[h]
\caption{}
\label{fig1}
A recursive construction of half-space trees.
\end{figure}

\begin{figure}[h]
\caption{}
\label{fig1bis}
$h_{i+1}^{ind}(h_i)$ where $h_i$ is
the total magnetic field
at generation $i$, and $h_{i+1}^{ind}$ is the
iterated field of equation
(\ref{eq17}). The three curves correspond to
(a) $\beta < \beta_c$, (b) $\beta = \beta_c$ and
(c) $\beta > \beta_c$.
\end{figure}

\begin{figure}[h]
\caption{}
\label{fig2}
The density probability of the magnetization on a
half-space tree where
$n =10$,$z=3$,
the ancestral  spin is fixed, and the inverse
temperature is $\beta=3$.
\end{figure}

\begin{figure}[h]
\caption{}
\label{fig23}
The conditional magnetization distribution in the
presence of a magnetic field,
the field being parallel to the central spin.
The magnetization distribution is plotted for a
 $z=3$ tree, $n=10$ generations,
and for magnetic fields $H=0,0.001,0.002$. The
magnetization distribution
evolves towards a gaussian shape as the magnetic
 field increases.
\end{figure}

\begin{figure}[h]
\caption{}
\label{fig3}
A typical configuration of the tree with
$n=8$ and $\beta=1.5$.
The vertices with no dots represent up spins
and the dots represent
flipped spins. Each kink gives rise to a
well-defined domain of flipped
spins.
\end{figure}

\begin{figure}[h]
\caption{}
\label{fig4}
Eigenvalues of the Glauber matrix as a function
of the inverse temperature
for an open six-site Ising chain (64 states).
The evolution of the energy levels as a function
of $\beta$ is monotonous,
and the eigenvalues are free to cross each other.
\end{figure}

\begin{figure}[h]
\caption{}
\label{fig5}
Spectrum of the Glauber matrix at a given
temperature for a ring of ten sites.
The inverse temperature is $\beta=3$. No gap is present.
\end{figure}

\begin{figure}[h]
\caption{}
\label{fig6}
Spectrum of the Glauber matrix at a given
temperature for a 3x3 Ising model
with open boundary conditions.
The inverse temperature is $\beta=3$.
Some degeneracies appear in the
spectrum, and the spectrum is symmetric
with respect to the $\lambda=-N/2$
line.
\end{figure}

\begin{figure}[h]
\caption{}
\label{fig7}
Eigenvalues of the Glauber matrix as a function
of the inverse temperature
for a 3x2 Ising cluster. The number of sites is 6,
leading to 64 states.
The evolution of the eigenvalues as a function of
$\beta$ is non monotonous,
and avoided crossings are visible.
\end{figure}

\begin{figure}[h]
\caption{}
\label{fig8}
Eigenvalues of the Glauber matrix as a function
of the inverse temperature
for a $n=1$, $z=3$ Cayley tree.
The number of sites is four, leading to 16 states.
The evolution of the energy
levels as a function of $\beta$ is not monotonous
for certain levels,
indicating level repulsion.
\end{figure}

\begin{figure}[h]
\caption{}
\label{fig9}
Spectrum of the Glauber matrix at a given
temperature for $z=3$ Cayley
tree with 2 generations. The inverse
temperature is $\beta=2$.
The spectrum is symmetric with respect
to the $\lambda=-N/2$ line.
The clusters collapse at integer values.
\end{figure}

\begin{figure}[h]
\caption{}
\label{fig10}
Evolution of the magnetization of the
16 pure states of the tree
with one generation. The inverse
temperature is $\beta=3$.
Metastable states appear to be present
even for such a small size.
The insert represents the same curve at
short time scales, which shows
the transient regime from the natural basis
of pure states to the metastable
states.
\end{figure}

\begin{figure}[h]
\caption{}
\label{fig11}
Expectation value of the central spin in the
 simplified dynamics, as
a function of the inverse temperature. The
transition temperature
is in agreement with the value
$\beta_C \simeq 0.54$.
The shape of the curve near the
transition is in agreement with the
existence of a mean field like
transition for the central spin.
We also plotted the expectation
value of the magnetization of
the $m$ closest to the origin
slices of spins, in the
factorisation approximation, for
$m=3,4,5,6,7,8,9,10$.
\end{figure}

\begin{figure}[h]
\caption{}
\label{fig12}
Magnetization relaxation of a single kink.
Initially, all the spins of
a $z=3$, $n=10$ tree are up, and one creates
a kink. We call $n_0$
the number of generations involved in the kink.
If $n_0=1$, the kink has only one spin, if
$n_0=2$, the kink has 7 spins,
etc. We follow the magnetization of this domain
as a function of time.
The unit time is one Monte Carlo Step (M.C.S.).
One M.C.S. corresponds
to repeating $N$ times the process which
consists in choosing one spin
at random among the $N$ sites, and changing
or not its direction,
according to the Boltzman distribution.
The curves are averaged over
50 different Monte Carlo runs of the dynamics.
\end{figure}

\begin{figure}[h]
\caption{}
\label{fig13}
Logarithm of the relaxation time as a function
of the number of generations
in the kink. The relaxation time is defined
from the cancelation of
magnetization on figure \ref{fig12}. The
points are approximately aligned,
which is in agreement with the Arrhenius
law of equation (\ref{eq13}).
\end{figure}

\begin{figure}[h]
\caption{}
\label{fig14}
$\chi_1'$ susceptibility on the Cayley tree.
The tree has 10 generations.
The amplitude of the magnetic field is $0.1$.
The curves are averaged over
50 initial configurations of spins, generated
at equilibrium.
One has to find a compromise between the
amplitude of the magnetic field and the
number of configurations to be averaged over,
to have a good signal/noise
ratio. The curves correspond to a
period of the magnetic field equal to 500
M.C.S., 1000 M.C.S. and 1500 M.C.S..
\end{figure}

\begin{figure}[h]
\caption{}
\label{fig15}
$\chi_3'$ susceptibility on the Cayley tree.
The tree has 10 generations.
The amplitude of the magnetic field is $0.1$.
The curves are averaged over
50 initial configurations of spins, generated
at equilibrium.
The curves correspond to a
period of the magnetic field equal to 500 M.C.S.,
1000 M.C.S. and 1500 M.C.S..
\end{figure}

\clearpage

\begin{figure}[h]
\caption{}
\label{fig17}
Spin autocorrelation functions of the
$z=3$ Cayley tree above $T_g$.
The unit time is one M.C.S., the inverse
temperature is $\beta = 0.5$, and
the averages are taken over $100$ random
initial configurations.
The autocorrelation functions decrease
rapidly with $t$ and are independant
on $t_w$.
\end{figure}

\begin{figure}[h]
\caption{}
\label{fig18}
Spin autocorrelation functions of the $z=3$
 Cayley tree below $T_g$.
The unit time is one M.C.S., the inverse
temperature is $\beta = 2.$, and
the averages are taken over $100$ random
initial configurations.
The autocorrelation functions depend on
the waiting time, and
{\it increase} with the waiting time,
 which is the signature of glassiness.
\end{figure}

\begin{figure}[h]
\caption{}
\label{fig19}
Edwards-Anderson order parameter for the whole
 spin system. The coordination is $z=3$.
The Edwards-Anderson order parameter is plotted
as a function of
temperature for $n=10$ (2047 spins) and $n=80$
(4.02 moles of spins).
The two curves nearly coincide.
\end{figure}

\begin{figure}[h]
\caption{}
\label{fig20}
Edwards-Anderson susceptibility for the whole
spin system. The coordination is $z=3$.
The Edwards-Anderson susceptibility is
plotted as a function of
temperature for $n=10$ (2047 spins) and
$n=80$ (4.02 moles of spins).
The two curves nearly coincide.
\end{figure}

\begin{figure}[h]
\caption{}
\label{fig21}
Edwards-Anderson susceptibility for the
central spin as a function of
temperature for $n=5,10,15,20,25,30,35,40$ generations.
The coordination is $z=3$.
\end{figure}

\begin{figure}[h]
\caption{}
\label{fig22}
Edwards-Anderson susceptibility for the central spin
as a function of
temperature for $z=3,4,5$. The number of
generations is $n=20$.
\end{figure}

\begin{figure}[h]
\caption{}
\label{fig25}
Edwards-Anderson susceptibility of the central
 spin in a magnetic field as a function of temperature,
for different sizes. The number of generations
 is n=2, 3, 4, 5, 6, 7, 8,
9, 10, 30.
The magnetic field is $H=0.05$
\end{figure}

\clearpage

\begin{figure}[h]
\caption{}
\label{fig26}
Maxima of $\chi_{EA}$ as a function of temperature for
different values of the magnetic
field. The system size is $n=10$ generations.
The magnetic fields are
$H=0.025, H=0.05, H=0.1, H=0.15, H=0.2$.
\end{figure}

\begin{figure}[h]
\caption{}
\label{fig27}
Zeros of Lee and Yang at different temperatures.
The radial coordinate is
rescaled in order to allow the superposition of
 different maps of zeros.
The external map represents the zeros for
12 generations, $z=3$ and
$\beta = 0.3$. In this case, the zeros are
localized on well defined
areas of the complex plane, far from the
real axis. The intermediate map
represents the case
$\beta = \beta_c \simeq 0.54$. The zeros begin to
fill the circle, with areas of zero density,
 espacially at the intersection
of the real axis with the unit circle.
The inner map represent the zeros
at low temperature, $\beta = 2$. In this case,
the circle is filled with
zeros. An analysis of the density of zeros
reveals an annormally high
density in the vicinity of the point $h=0$.
\end{figure}

\begin{figure}[h]
\caption{}
\label{fig28}
Ratio of the density of zeros in the
vicinity of the point
$\exp{2 \beta h}=0$ over the average
density of zeros in the case of
the $z=3$ tree with 17 generations.
The inset represents the same
quantity for a Weiss model with $T_c=1$.
In the case of the tree, the ration is
greater than one below a certain
temperature, whereas it is
always less than unity for the Weiss model.
\end{figure}


\newpage
\begin{thebibliography}{99}

\bibitem{thorpe} For a pedagogical review of recursive structures see
M.F. Thorpe in {\sl Excitations in Disordered Solids (NATO Advanced
Study Institute Series B}, ed. M.F. Thorpe (New York: Plenum, 1982),
pp 85-107.

\bibitem{domb}C. Domb, {\sl Adv. Phys.}, {\bf 9}, 145 (1960).

\bibitem{baxter}e.g. R.J. Baxter, {\sl Exactly Solved Models in Statistical
Mechanics}, (Academic Press, London, 1982), Chapter 4.

\bibitem{gujrati}P.D. Gujrati, {\sl Phys. Rev. Lett.}, {\bf 74}, 809 (1995).

\bibitem{tap}D.J. Thouless, P.W. Anderson and R.G. Palmer, {\sl Phil.Mag.},
{\bf 35}, 593 (1977).

\bibitem{sk} D. Sherrington and S. Kirkpatrick, {\sl Phys. Rev. Lett.},
{\bf 35}, 1792 (1975).

\bibitem{1}D. Bowman and K. Levin, {\sl Phys. Rev. B}, {\bf 25}, 3438 (1982).

\bibitem{2}D.J. Thouless, {\sl Phys. Rev. Lett.}, {\bf 56}, 1082, (1986).

\bibitem{3}S. Katsura, {\sl Prog. Theo. Phys.}, {\bf 87}, 139 (1986).

\bibitem{4}J.T. Chayes, L. Chayes, J.P. Sethna and D.J. Thouless,
{\sl Comm. Math. Phys.}, {\bf 106}, 41 (1986).

\bibitem{5}P. Mottishaw, {\sl Europhys. Lett.}, {\bf 4}, 333 (1987).

\bibitem{6}Pik-Uin Lai and Y.Y. Goldschmidt, {\sl J. Phys. A}, {\bf 22}, 399
(1989).

\bibitem{7}J.M. Carlson, J.T. Chayes, L. Chayes, J.P. Sethna and
D.J. Thouless, {\sl Europhys. Lett.}, {\bf 5}, 355 (1988);
{\sl J. Stat. Phys.}, {\bf 61}, 987 (1990).

\bibitem{8}J.M. Carlson, J.T. Chayes, J.P. Sethna and D.J. Thouless,
{\sl J. Stat. Phys.}, {\bf 61}, 1069 (1990).


\bibitem{9}Y.Y. Goldschmidt, {\sl Phys. Rev. B}, {\bf 43}. 8148 (1991).

\bibitem{polymers}P.J. Gujrati, {\sl Phys. Rev. Lett.}, {\bf 53}, 2453 (1984);
{\sl J. Chem. Phys.}, {\bf 98}, 1613 (1993).

\bibitem{stinchcombe}R.B. Stinchcombe, {\sl J. Phys. C}, {\bf 6}, L1 (1973).

\bibitem{rammal}R. Rammal, {\sl J. Physique}, {\bf 46}, 1837 (1985);
R. Rammal and A. Benoit, {\sl J. Physique Lett.}, {\bf 46}, L-667 (1985);
{\sl Phys. Rev. Lett.}, {\bf 55}, 649 (1985).

\bibitem{10}T.P. Eggarter, {\sl Phys. Rev. B}, {\bf 9}, 2989 (1974).

\bibitem{11}M. Matsuda, {\sl Prog. Theor. Phys.}, {\bf 51}, 1053
(1974).

\bibitem{12}J. von Heimburg and H. Thomas, {\sl J. Phys. C}, {\bf 7},
3433 (1974).

\bibitem{13}E. Muller-Hartmann and J. Zittartz, {\sl Phys. Rev.
Lett.}, {\bf 33}, 893 (1974).

\bibitem{ref3} R.J. Glauber, Jour. Math. Phys.
\underline{4}, 294 (1963).

\bibitem{ref4} L.P. L\'evy, Phys. Rev. B
\underline{38}, 4963 (1988).

\bibitem{ref5} E. Marinari, G. Parisi and
F. Ritort, preprint
cond-mat/9410089.

\bibitem{ref6} M. Kurata, R. Kikuchi and T. Watari, J. Chem Phys,
\underline{21}, 434 (1953).

\bibitem{ref8} C.N. Yang, T.D. Lee, Phys. Rev.
\underline{87}, 404 (1952),
T.D. Lee, C.N. Yang, Phys. Rev. \underline{87},
 410 (1952). See also
K. Huang, {\it Statistical Mechanics}, Wiley, N.Y. (1963).

\bibitem{ref9} R. Rammal, th\`ese d'\'etat (1981).

\bibitem{aps} J. C. Angl\`es d'Auriac, M. Preissmann and
A. Seb\H{o} submitted to Journal of Mathematics and Combinatorics (1995).
\end{thebibliography}
\end{document}